\documentclass[
    ,final            
  ]
  {aipproc}

\layoutstyle{6x9}

\begin{document}

\title[The clustering of quasars]{The clustering of simulated quasars}


\author{Silvia Bonoli}{
  address={Max-Planck-Institut f\"ur Astrophysik, Karl-Schwarzschild Strasse 1,
D-85740 Garching, Germany}, 
, email={bonoli@mpa-garching.mpg.de}}

\begin{abstract}
We analyze the clustering properties of quasars simulated using a
semianalytic model built on the Millennium Simulation, with the goal of testing
scenarios in which black hole accretion and quasar activity are triggered by
galaxy mergers. When we select quasars with luminosities in the range
accessible by current observations, we find that predicted values for the
redshift evolution of the quasar bias agree rather well with the available data
and the clustering strength depends only weakly on
luminosity. This is independent of the lightcurve model assumed, since bright quasars are black
holes accreting close to the Eddington limit.  
We also used the large catalogues of haloes  available for the
Millennium Simulation to test whether recently merged haloes exhibit a stronger
large-scale clustering than the typical haloes of the same mass. This effect
might help to explain  the very high clustering
strength observed for $z\sim 4$ quasars. However, we do not detect any significant
 excess bias for the clustering of merger remnants, suggesting 
that objects of merger-driven nature do not cluster significantly differently
than other objects of the same characteristic mass.
\end{abstract}

\maketitle


\section{INTRODUCTION}

In recent years, the analysis of quasar clustering has proven to be an
important tool for understanding  not only the environment, but also the properties,
such as lifetimes, of these objects. If quasars exhibit a strong clustering,
they  must be hosted by
rare and massive dark matter haloes and thus they would
have to be long events in order to account for the total observed quasar luminosity
density. A weak clustering would instead indicate more numerous and less massive
hosts; in this case  many short accretion events would have to contribute to the total 
luminosity output \citep{cole89, martini01,
haiman01}. 

Wide-field surveys like the Sloan Digital Sky Survey
(SDSS) and the 2dF quasi-stellar object (2dFQSO) survey have been able to observe
thousands of quasars up to $z\sim5$, allowing a
detailed investigation of their clustering properties and their redshift
evolution
\citep[e.g.,][]{porciani04, croom05, shen07}. The quasar clustering strength has
been observed to be an increasing function of redshift, with an 
evolution consistent with the one of dark matter haloes. 
Assuming that the  clustering strength of haloes  depends only on
their mass, these studies concluded that quasars reside at all times in haloes
with mass
 $M_{\rm halo} \sim 3\times 10^{12} -  10^{13} h^{-1} \rm{M}_{\odot}$. 
The corresponding quasar lifetime would be a few times $10^{7}$ yr, reaching $10^{8}$
yr at the highest observed redshifts.
The same observations have also indicated that quasar clustering does not
significantly depend on luminosity. However, this result might be partly explained by the
narrow range of  magnitudes covered by these surveys. When  \citet{shen09}
analyzed the  brightest $10 \%$ of objects in their sample, they found that these
quasars have a higher bias compared to the full sample.

Interestingly, the very high clustering amplitude of luminous quasars at
$z>3$ measured by \citet{shen07} has posed some 
theoretical problems for the simultaneous interpretation of the
clustering and the luminosity function at these epochs. The high clustering
would suggest that these quasars live in very massive haloes, but the extreme
rareness of such haloes is difficult to
reconcile with the observed quasar number density and luminosity function, especially at
$z\sim 4$. Various theoretical works have suggested that clustering and number
densities can be matched only by assuming a high quasar duty cycle, a very low scatter in the
 relation between halo mass and quasar luminosity \citep{White08} and high
radiative efficiencies  \citep{ShankarCrocce}. Recently,
\citet{wyithe09} suggested that these conclusions can be relaxed if haloes
hosting quasars cluster more strongly than typical haloes of similar mass. This would then
imply that quasars live in less massive but more numerous haloes,
allowing for lower duty cycles and less extreme values for the radiative
efficiency. In
particular, \citep{wyithe09} suggests that the possible merger-driven
nature of quasars might cause an excess bias, if the large-scale
clustering of recently-merged haloes is higher than 
expected for typical haloes of the same mass (``merger bias''). 

In what follows, we first show a comparison between the clustering of
simulated quasars with the most recent observational data to test our model of
black holes accretion. We then explore the possibility that recently merged haloes are
more clustered than other haloes of the same mass and discuss the implications
that such an effect could have in the interpretation of quasar clustering.

\section{The clustering of simulated quasars}
\label{sec:clustering}

\begin{figure}
  \includegraphics[height=.28\textheight]{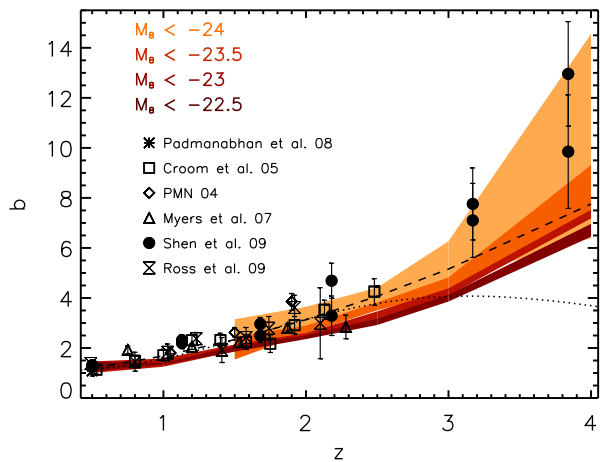}
  \caption{Bias of simulated quasars selected using four B-Band magnitude cuts
as indicated on the plot (colored regions, whose width is given by the
$1-\sigma$ uncertainty), compared with observational data at
various redshifts. The dotted line is the prediction of \citet{hopkins07b} and
the short-dashed line is the best fit from \citet{croom05}.}
   \label{fig:qso_bias}
\end{figure}

We model the evolution of supermassive black holes (BHs) and quasars using 
 the semianalytic model for galaxy formation built on the Millennium
Simulation \citep[e.g.,][]{delucia07}. We assume that every newly-formed galaxy hosts a central BH of $10^3
\rm{M}_{\odot}$, which grows by efficiently accreting gas during galaxy mergers,
according to the parametrization of \citet{croton06b}. The bolometric luminosity
associated with each accretion event is modeled as in \citet{marulli08}. Using the
bolometric conversions of \citet{hopkins07} we extracted subsamples of quasars
visible in the optical. We then calculate the bias $b$ of quasars with different
luminosities, where $b \equiv (\xi_{\rm QSO}/ \xi_{\rm DM})^{1/2}$ and
$\xi_{\rm QSO}$ and  $\xi_{\rm DM}$ are the two-point autocorrelation function of
the quasars and the dark matter, respectively.  In Fig. \ref{fig:qso_bias} the
bias of  quasars with different luminosities is shown as a function
of redshift and compared with recent observational data. The prediction for the
redshift evolution of quasar bias is
independent of the assumed lightcurve model of single accretion events, because
such bright quasars are BHs accreting close to the Eddington limit. 

The good agreement with
 the data indicates that our merger-triggered BH accretion model predicts a
spatial  distribution of quasars
that is consistent with observations \citep[for
further details, see][]{bonoli09}. This non-trivial outcome can be
viewed as a further success of the hierarchical galaxy formation paradigm, since
the location of halo mergers is a direct prediction of dark matter cosmological simulations.

\section{The merger bias}
\label{sec:merger_bias}

\begin{figure}
  \includegraphics[height=.47\textheight]{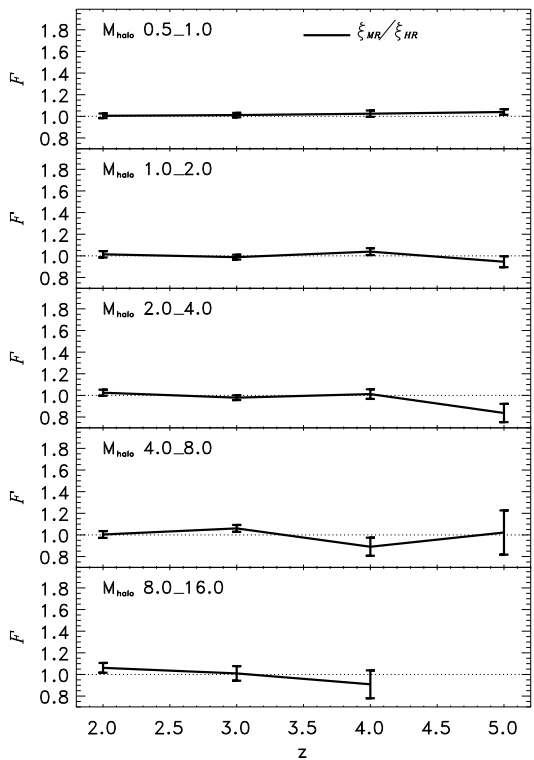}
  \caption{Excess bias for DM haloes in separate mass bins, as
          indicated in each panel in units of $10^{12} \, h^{-1} \,
          \rm{M}_{\odot}$. The horizontal dotted line corresponds to an absence of
	  excess bias (i.e., $\mathit{F}=1$). }
  \label{fig:F_haloes}
\end{figure}

We use the large catalogues of haloes available for the Millennium
Simulation to quantify the importance of merger bias. If statistically
significant, this effect could
 help to understand the very strong clustering observed for high-redshift quasars. 
We define as major mergers those events in which two haloes of
comparable mass merge into a single system between two outputs of the
simulation. We then compare the clustering of merger remnants of a
given mass with the clustering  of all haloes with the same mass. Figure
\ref{fig:F_haloes} shows, for different mass bins, the redshift evolution  of the
 excess bias $\mathit{F}$, defined as the ratio of the two point correlation
function of the merger remnants and the two point correlation function of the
entire halo population with the same mass. 
Clearly, at all redshifts and for all halo
masses, no significant excess bias is present. This result indicates that haloes
that recently merged have  large-scale clustering properties that do not
significantly differ from the clustering properties of other haloes of the same
mass. We also looked for a possible merger bias among samples of galaxies
selected from the semianalytical model of galaxy formation mentioned above. For
the galaxies we find a more significant merger bias ($F\sim 1.2$), which,
however, decreases with
increasing stellar mass \citep[for further details, see][]{bonoli09b}.  

The weak merger bias of massive systems suggests that objects of merger-driven
nature, such as bright quasars, do not cluster significantly differently than other objects of
the same characteristic mass. We do find that, if quasars are triggered by
 major merger events, their clustering and number
densities can be reconciled if we adopt high duty cycles and small scatter in
the relation between quasar luminosity and halo mass.


\begin{theacknowledgments}
I am grateful to Enzo Branchini, Federico Marulli, Lauro Moscardini, Francesco Shankar, Volker Springel,
Simon White and Stuart Wyithe for their essential contribution to the work presented here.
\end{theacknowledgments}



\bibliographystyle{aipproc}   

\bibliography{bonoli_s}

\IfFileExists{\jobname.bbl}{}
 {\typeout{}
  \typeout{******************************************}
  \typeout{** Please run "bibtex \jobname" to optain}
  \typeout{** the bibliography and then re-run LaTeX}
  \typeout{** twice to fix the references!}
  \typeout{******************************************}
  \typeout{}
 }

\end{document}